# An efficient source of frequency anti-correlated entanglement at telecom wavelength

Feiyan Hou[1,2] · Xiao Xiang[1,2] · Runai Quan[1,2] · Mengmeng Wang[1,2] · Yiwei Zhai[1,2] · Shaofeng Wang[1,2] · Tao Liu[1] · Shougang Zhang[1] · Ruifang Dong[1]

**Abstract** We demonstrate an efficient generation of frequency anti-correlated entangled photon pairs at telecom wavelength. The fundamental laser is a continuous-wave high-power fiber laser at 1560 nm, through an extracavity frequency doubling system, a 780-nm pump with a power as high as 742 mW is realized. After single-passing through a periodically poled KTiOPO4 (PPKTP) crystal, degenerate down-converted photon pairs are generated. With an overall detection efficiency of 14.8 %, the count rates of the single photons and coincidence of the photon pairs are measured to be 370 kHz and 22 kHz, respectively. The spectra of the signal and idler photons are centered at 1560.23 and 1560.04 nm, while their 3-dB bandwidths being 3.22 nm both. The joint spectrum of the photon pair is observed to be frequency anti-correlated and have a spectral bandwidth of 0.52 nm. According to the ratio of the single-photon spectral bandwidth to the joint spectral bandwidth of the photon pairs, the degree of frequency entanglement is quantified to be 6.19. Based on a Hong–Ou–Mandel interferometric coincidence measurement, a frequency indistinguishability of 95 % is demonstrated. The good agreements with the theoretical estimations show that the inherent extra intensity noise in fiber lasers has little influence on frequency entanglement of the generated photon pairs.

## 1 Introduction

Entangled photon pairs based on spontaneous parametric down-conversion (SPDC) process [1] are not only at the heart of the most fundamental tests of quantum mechanics [2, 3] but also essential resources for quantum information processing (QIP) [4]. The entanglement between the generated signal and idler photons can be in polarization, photon quadrature, angular momentum, frequency, wave vector, etc. Recently, frequency entanglement have attracted many interests due to its wide applications in quantum enhanced positioning and clock synchronization [5–14], quantum spectroscopy [15], quantum optical coherence tomography [16–20], quantum imaging [21, 22], nonlinear microscopy [23–25], and quantum key distribution [26, 27].

Because of low-loss, high-stability features of optical fibers, they become a widely used transmission medium. In a fiber-based quantum information network, frequency-entangled photon pair source at telecom wavelength is required for minimum absorption. Several experiments on generation of frequency-correlated photon pairs at telecom wavelength have been reported [28–31] since pulse pumped SPDC can be generated and detected much more efficiently. However, in application to the entanglement-based quantum communication, their broad spectral width of the photon pairs limits the fiber transmission distance. By using of frequency anti-correlated photons, the deleterious effects of chromatic dispersion can be mitigated. Particularly, frequency anti-correlated photon pairs have unprecedented advantage of dispersion cancelation in quantum synchronization [6, 11, 14]. Therefore, the generation of frequency anti-correlated entangled photon pairs in the telecom wavelength regime and with a high brightness is required for the long-distance quantum communication.

✉ Ruifang Dong
dongruifang@ntsc.ac.cn

[1] Key Laboratory of Time and Frequency Primary Standards, National Time Service Center, Chinese Academy of Science, Xi'an 710600, China

[2] University of Chinese Academy of Sciences, Beijing 100049, China

Springer

**Fig. 1** The experimental setup for generation and characterization of the frequency anti-correlated entangled biphoton source at 1560 nm. The *inset* labeled by *A* denotes the spectral measurement setup, while *B* is for the HOM interferometric measurement

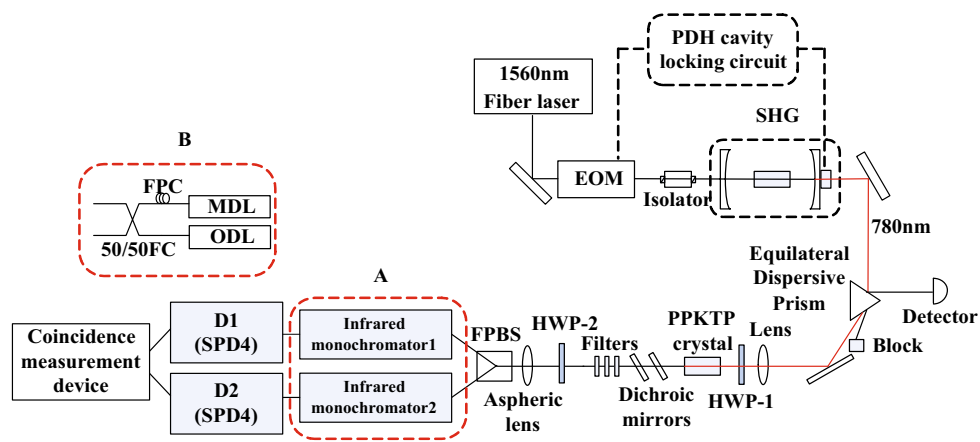

For the generation of frequency anti-correlated entangled photon pairs at 1560 nm, a continuous-wave (cw) 780-nm laser source is required as the pump of a type-II SPDC. However, most of such 780-nm laser sources with a power higher than 200 mW are acquired by a Ti:Sapphire laser pump, the application of which is limited by their high cost. On the other hand, with the rapid development of fiber amplifiers, a low-cost, high-power, single-frequency fiber laser at 1560 nm can be obtained based on a $Er^{3+}/Yb^{3+}$-doped single-mode phosphate glass fiber [32, 33]. For acquiring a high-power cw 780-nm laser source, frequency doubling of a high-power 1560-nm fiber laser becomes a preferable method [32, 34].

In this paper, we demonstrate an efficient generation of frequency anti-correlated entangled photon pairs at 1560 nm. First the generation of a high-power cw laser source at 780 nm with an extracavity frequency doubling of a custom-made 1560-nm single-frequency fiber laser is presented,[1] the power of which can achieve 742 mW. Based on this source and a type-II SPDC crystal of PPKTP, frequency anti-correlated entangled photon pairs at 1560 nm are then generated. With an overall detection efficiency of 14.8 %, the count rates of the single photons and coincidence of the photon pairs are measured to be 370 kHz and 22 kHz respectively. Compared with the first report on such photon pairs generation [41], the single-count rate has been improved for sevenfold while the coincidence-count rate for fourfold. According to the spectroscopic measurements, the spectra of the signal and idler photons are centered at 1560.23 and 1560.04 nm while their 3-dB bandwidths being 3.22 nm both, while the joint spectrum of the photon pair is observed to be frequency anti-correlated and have a spectral bandwidth of 0.52 nm. Thus the degree of frequency entanglement is quantified to be 6.19. Based on a Hong–Ou–Mandel (HOM) interferometric coincidence measurement, a frequency indistinguishability of 95 % is demonstrated. The good agreements with the theoretical estimations show that the inherent extra intensity noise in fiber lasers has little influence on frequency entanglement of the generated photon pairs.

The rest of the paper is organized as follows. The experimental setup for the generation of the frequency anti-correlated entangled state at 1560 nm is described in Sect. 2, and a theoretical model of the setup is presented in Sect. 3. Experimental results are presented in Sect. 4. Finally, Sect. 5 contains a brief conclusion.

## 2 Experimental setup

The setup used for experimental generation of frequency anti-correlated entangled biphoton source at 1560 nm can be divided into two parts. As shown in Fig. 1, the first part is for generation of the frequency-doubling-based quasi-monochromatic 780-nm pump source. The fundamental laser source we use is a single-frequency 1560-nm fiber laser with a linewidth of 2 kHz and a maximum output power of 2.5 W (see footnote 1). The frequency doubling external cavity is composed of two concave mirrors with an identical radius of 50 mm, and the cavity length is set to be 103.5 mm. The incoupler M1 has a high reflectivity of R1 > 99 % for the second harmonic 780 nm beam, while the transmittance to 1560 nm is 5 %. The outcoupler M2 is 99.9 % reflective to 1560 nm and its transmittance to 780 nm is 95.2 %. A piezo-actuator (PZT) is attached to M2 for scanning and locking the cavity length. To realize the second harmonic generation, a type-I PPKTP crystal (Raicol Ltd.) with a size of 1 mm × 2 mm × 10 mm and a poling period of 24.925 µm is inserted into the center of the cavity. The working temperature is optimized and fixed at 70°C for maximum output harmonic power. Before coupling into the second harmonic cavity, the fundamental 1560-nm fiber laser passes first through an EO modulator and an optical isolator and then is collimated to focus onto the center of PPKTP with a beam waist of 66 µm. The total transmittance out of these optics is measured to be 64.2 %. Applying PDH locking technique [35], a stable second harmonic generation at 780 nm can be output and used as the pump of the subsequent SPDC.

---

[1] The laser is provided by Prof. Z. Yangs group of South China University of Technology, which is a collaborative product of the Cross and Cooperative Science and Technology Innovation Team Project of the CAS, China.

After filtering out the residual fundamental 1560-nm laser by an equilateral dispersive prism and a series of short-pass filters (FESH1000, Thorlabs), the generated 780-nm laser subsequently single passes through a type-II PPKTP crystal for generating the down-converted signal and idler photons with orthogonal polarizations. The PPKTP has a length of 10 mm and a poling period of 46.146 µm, while the 780-nm beam is focused to a waist of 48 µm in the center of PPKTP. Through careful evaluation, the temperature of PPKTP is set at 64°C to ensure frequency degeneracy of the generated photon pairs. Via a series of dichroic mirrors and long-pass filters, the 780-nm laser is filtered out from the generated photon pairs. Afterward, the photon pairs are coupled into a fiber polarization beam splitter (FPBS). With the help of a half-wave-plate (HWP) at the entrance, the signal and idler are separated into two output ports of FPBS. Linking the two ports to two infrared monochromators (MONO 1&2, Jobin Yvon MicroHR, see inset A of Fig. 1), the joint spectral properties of the photon pairs can be measured [36] by scanning the MONOs and monitoring the outputs of them with the help of two single photon detectors (SPD4, D1 & D2) and subsequent coincidence measurement device (Ortec 453). When we measure the individual spectrum of the photon pairs, only one of the MONOs is connected, while the other output of the FPBS is sent directly to the single photon detector. With the information of the individual and joint spectral properties of the down-converted photon pairs, the degree of frequency entanglement can then be evaluated. Furthermore, replacing the MONOs setup with a fiber-based HOM interferometer [37] as shown in inset B, the frequency indistinguishability can be measured. In the HOM interferometer shown in inset B, the ODL is a manually adjusted optical delay line (ODL, General Photonics Inc.), while the MDL is a motorized optical delay line (MDL-002, General Photonics Inc.) which is adjustable with a resolution of 1 fs within the range of 0–560 ps. A fiber polarization controller (FPC) is inserted in one arm to match the polarization in the other arm.

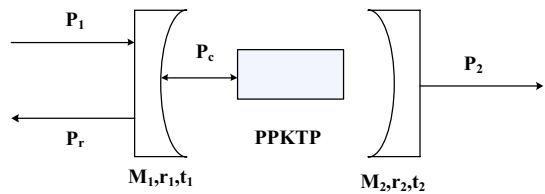

**Fig. 2** Theoretical model of a singly-resonant two-mirror SHG cavity

## 3 Theoretical model

### 3.1 Second harmonic generation of 780-nm source

In the experiment, we use a singly-resonant two-mirror standing wave cavity for second harmonic generation (SHG). Its theoretical model is shown in Fig. 2.

Assume that the reflectivity and transmittance of incoupler (outcoupler) $M_1(M_2)$ to the fundamental 1560-nm source are $r_1(r_2)$ and $t_1(t_2)$ respectively. The transmittance of $M_2$ to 780 nm is $t_{2SH}$. $t$ denotes the single-pass efficiency of the fundamental source through the cavity, $t = 1 - \delta$, where $\delta$ represents the total intra-cavity loss. $\gamma_{SH}$ is the nonlinear conversion coefficient, which can be deducted according to experimental parameters [38]. The output power of the second harmonic generation is then given by

$$P_2 = 2\gamma_{SH} P_c^2 t_{2SH}, \tag{1}$$

where $P_c$ denotes the intra-cavity cycling power of the fundamental source. It is related to the input power by the following formula:

$$P_c = \frac{t_1}{(1 - \sqrt{r_1 r_m})^2} P_1, \tag{2}$$

where $r_m$ describes the total transmission efficiency of the fundamental source after a single round-trip in the SHG cavity, it is given by

$$r_m = t^2 \cdot (1 - \gamma_{SH} \cdot P_c)^2 \cdot r_2. \tag{3}$$

Combining the above formulae, the second harmonic power out of the SHG cavity as a function of the input fundamental power can be analyzed.

### 3.2 Characterization of the generated photon pairs

With the second harmonic generated 780-nm source as the pump, a type-II PPKTP crystal is used to generate frequency anti-correlated entangled photon pairs. The two-photon state can be expressed as [39, 40, 42, 43]

$$|\Psi\rangle = \iint d\omega_s d\omega_i A(\omega_s, \omega_i; T) a_s^\dagger(\omega_s) a_i^\dagger(\omega_i) |0\rangle, \tag{4}$$

where $\omega_s$, $\omega_i$ denote the frequencies of the signal and idler photon, respectively; $A(\omega_s, \omega_i; T)$ denotes the joint spectral amplitude function of the generated photon pairs, which is given by the product of the pump spectrum $\alpha(\omega_s, \omega_i)$ and the phase-matching function $\Phi_L(\omega_s, \omega_i; T)$ of the nonlinear crystal.

$$A(\omega_s, \omega_i; T) = \alpha(\omega_s, \omega_i)\Phi_L(\omega_s, \omega_i; T). \tag{5}$$

The spectrum of the pump is written as

$$\alpha(\omega_s, \omega_i) \propto \exp\left[-\frac{(\omega_s + \omega_i - \omega_p^0)^2}{4B_p^2}\right], \tag{6}$$

where $\omega_p^0$ and $B_p$ denote the center frequency and spectral bandwidth of the pump. The phase-matching function satisfied by the SPDC photon pairs is given by

$$\Phi_L(\omega_s, \omega_i; T) \equiv \frac{\sin[\Delta k(\omega_s, \omega_i; T)L/2]}{\Delta k(\omega_s, \omega_i; T)/2},$$
$$\Delta k(\omega_s, \omega_i; T) \equiv k_p(\omega_p; T) - k_s(\omega_s; T) - k_i(\omega_i; T) \pm 2\pi/\Lambda. \tag{7}$$

where $L$ is the crystal length, $T$ and $\Lambda$ represent the working temperature and poled period of PPKTP, respectively. $\Delta k$ is the wavevector mismatching. $k_j(\omega_j; T) = n(\omega_j; T)\omega_j/c, j = p, s, i$ denote the propagation constants of the pump, signal and idler respectively. Let us consider the case of collinear degenerate down-conversion, i.e., $\omega_s^0 = \omega_i^0 = \omega_p^0/2$, the deviations from the central frequencies are given by $\Omega_{s,i} = \omega_{s,i} - \omega_p^0/2$. Without loss of generality, $\Delta k$ can be Taylor expanded to its first-order terms

$$\begin{aligned}\Delta k(\omega_s, \omega_i) \equiv &(k_p^0(T) - k_s^0(T) \\ &- k_i^0(T) \pm 2\pi/\Lambda) \\ &- (k_p'(\omega_p^0; T) - k_s'(\omega_p^0/2; T))\Omega_s \\ &- (k_p'(\omega_p^0/2; T) - k_s'(\omega_p^0/2; T))\Omega_i,\end{aligned} \tag{8}$$

Based on the Sellmeier equation of PPKTP and its temperature dependence given in [44–46], the crystal temperature can be optimized to achieve the phase-matching condition $(k_p^0(T) - k_s^0(T) - k_i^0(T)) \pm 2\pi/\Lambda = 0$.

To quantify a frequency-entangled source, two figures of merit are used, which are the frequency indistinguishability and the degree of frequency entanglement. The frequency indistinguishability refers to the similarity between the spectral distributions of the down-converted signal and idler photons, which can be readily measured by the visibility of the HOM interferometric measurement [39]. In the theoretical form, it is given by the relative overlap function

$$V = \frac{\iint d\omega_s d\omega_i |A(\omega_s, \omega_i) A(\omega_i, \omega_s)|}{\iint d\omega_s d\omega_i |A(\omega_s, \omega_i)|^2}. \tag{9}$$

Alternatively, it can be measured by the overlap between the individual spectra of signal and idler photons [29].

The degree of frequency entanglement denotes the non-classical correlation between the spectral distributions of the signal and idler, which is fundamentally characterized by the Schmidt number $K$ [47]. The larger $K$ is, the higher the entanglement. Since the Schmidt number $K$ cannot be measured directly, an operational entanglement parameter $R$ is introduced [48, 49], which is given by

$$R = \Delta\omega_s/\delta\omega_c, \tag{10}$$

where $\Delta\omega_s$ and $\delta\omega_c$ represent the single-particle and coincidence spectral bandwidths, respectively. The coincidence photon spectrum can be given by the joint spectral density function $|A(\omega_s, \omega_i)|^2$ at a given value of $\omega_i$, e.g., $S_c(\omega_s) = |A(\omega_s, \omega_i = \omega_p^0/2)|^2$. The single-particle photon spectrum can be determined by $|A(\omega_s, \omega_i)|^2$ integrated over $\omega_i$, i.e., $S_s(\omega_s) = \int d\omega_i |A(\omega_s, \omega_i)|^2$.

In the experiment, the pump radiation is centered at wavelength of 780 nm and the 3-dB bandwidth is measured to be around 0.05 nm. In the phase-matching condition, the pump spectrum function and the phase-matching function based on the wavelength of signal and idler photons are shown in Fig. 3a, b. The joint spectrum of the photon pairs determined by the pump spectrum function and the phase-matching function is shown in Fig. 3c. The spectra of both signal and idler are shown in Fig. 3d.

According to the theoretical computation, the center wavelength of signal and idler photon are 1560.0 and 1559.9 nm, respectively. And the bandwidth of signal and idler photon are both 2.4 nm. The coincidence width of the joint spectrum is 0.43 nm. According to the criteria of frequency entanglement introduced in [48], the degree of frequency entanglement was determined to be $R = \Delta\lambda_s/\delta\lambda_c = 5.58$. The HOM visibility is estimated to be 99.9 % with a FWHM dip width of 1.48 ps.

## 4 Experimental results

### 4.1 Generation of the 780-nm pump source

By applying PDH locking to the SHG cavity, the generated second harmonic laser at 780 nm as a function of the fundamental laser power is experimentally investigated and shown in Fig. 4. At an incident fundamental power of 1.41 W (the power of 1560 nm laser is attenuated by the optical components, such as EO modulator, optical isolator, for about 64.2 % before it arrives at the SHG cavity), a power as high as 742 mW of 780-nm laser is generated and the corresponding second harmonic generation efficiency reaches 52.6 %. The theoretical simulation is also shown in Fig. 4a by black line. It can be seen that, the experimental

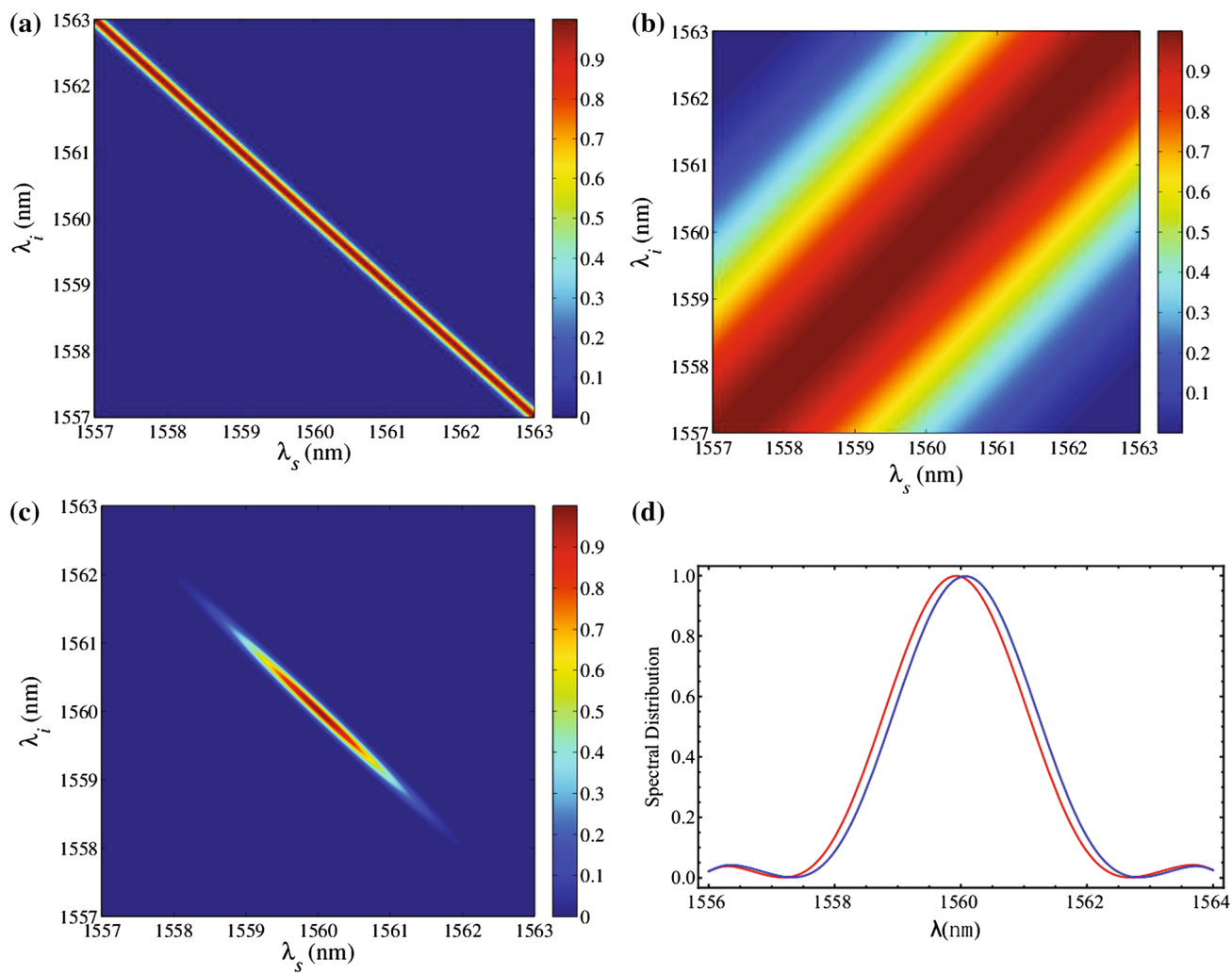

**Fig. 3** Figure in theory. **a** the pump spectrum function, **b** the phase-matching function, **c** the joint spectrum of the photon pairs, **d** spectra of both signal and idler photon

result fits well with the theoretical curve when the power of the fundamental laser is low. While further increasing the power of the fundamental laser, the difference between the experimental and theoretical results becomes obvious. It can be explained by transform of the generated 780-nm laser into 1560 nm when the power in the cavity is above the threshold of the parameter down-conversion [50]. The long-term stability of 780-nm output is monitored. The second harmonic cavity can be locked stably for as long as 70 h can be readily achieved and the result is given in Fig. 4b.

The intensity noise of the fundamental fiber laser always has a considerable excess intensity noise [51], it may make an effect on the generated the 780-nm laser and the properties of the subsequent frequency entanglement states. We first measure the intensity noise spectrum of the 780-nm laser by using a pair of home-made balanced homodyne detectors with a measurement bandwidth of 5 MHz. With 5 mW of optical power incident into each detector, the homodyne output is then analyzed by a spectrum analyzer (ROHDE&SCHWARZ FSH4) with a RBW of 30 kHz and a VBW of 300 Hz. As shown in Fig. 5, it has a rather high excess intensity noise, which remains 20 dB above the shot noise level until the analyzing frequency of 3 MHz.

### 4.2 Characterization of generated photon pairs at 1560 nm

With the above generated 780-nm laser as the pump and the type-II PPKTP crystal as the SPDC crystal, the down-converted signal and idler photons are then generated with orthogonal polarizations. As shown in Fig. 1, after filtering out the residual 780-nm laser by a serials of dichroic mirrors and filters, the frequency-entangled photon pairs

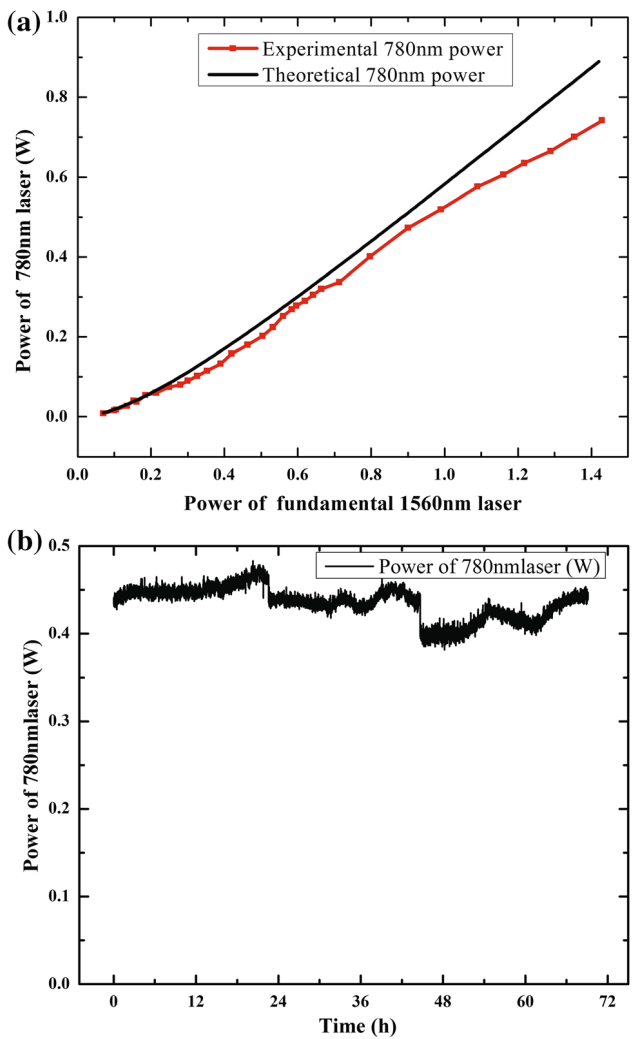

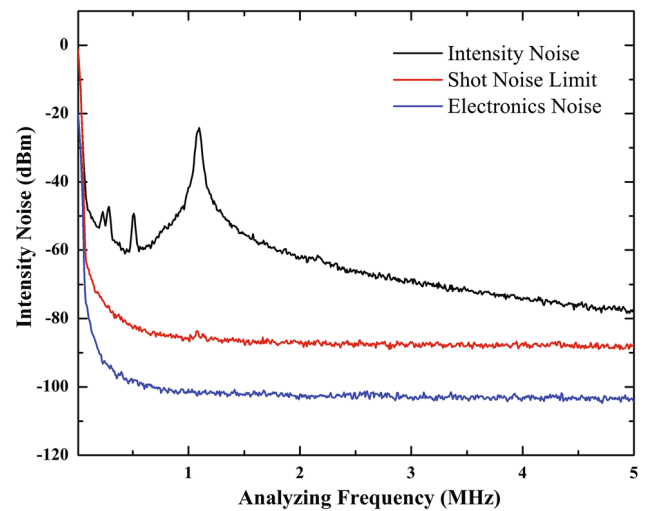

Fig. 5 The measured noise power of the generated 780-nm source (*black line*). The *red line* gives the relevant shot noise limit, and the *blue line* is the electronics noise of the detector. The measurement parameters: the power of 780-nm laser is 5 mW; the resolution bandwidth of the spectrum analyzer is 30 kHz while the video bandwidth is 300 Hz

Fig. 4 **a** The theoretical and experimental power of second harmonic laser at 780 nm as the function of the power of fundamental laser at 1560 nm. **b** The long-term stability of 780-nm output

are coupled into a FPBS. The fiber coupling efficiency is measured to be 74 %, by linking the two output ports of FPBS directly to two single photon detectors and subsequent coincidence measurement device, both the single photon count rate and the coincidence rate are measured. In our experiment, the two APD are used in the gated mode triggered by an external 75-MHz square-wave signal from a waveform generator (Tektronix AFG3252), the detection efficiency is 20 %. The rate of the singles were then measured to be 370 kHz, while the coincidence rate was around 22 kHz. The heralding efficiency $\mu_{s(i)}$ for the signal (idler) is defined to be the ratio between coincidences $C$ to the number of singles (detector noise subtracted) in the idler (signal) detector $S_{i(s)}$ corrected by the detection efficiency $\eta_{s(i)}$. In our experiment, the heralding efficiency of $\mu_{s(i)}$ is evaluated to be 40 %. The small pump waist $\omega_{p_0} = 48\,\mu$m, which corresponds to a focusing parameter of $\xi = 1.08$, should be the reason why such low heralding efficiency was achieved [52].

### 4.2.1 Spectral measurement of the photon pairs

Remaining only one of the monochromators in the setup (as shown in Fig. 1) while the other output of the FPBS being set to the single photon detector, the singles spectra of both signal and idler are shown in Fig. 6a. Through Gaussian fitting, we achieved the center wavelengths of the signal and idler to be $1560.23 \pm 0.03$ nm and $1560.04 \pm 0.03$ nm, while the 3-dB bandwidths of the signal and idler are both $3.22 \pm 0.01$ nm. We further measured the joint spectrum of the photon pairs, and the resultant two-photon spectral intensity is shown in Fig. 6b. The color bar on the right hand of the figure shows the corresponding coincidence counts for different colors. We can see that the joint spectrum has an anti-diagonal distribution. The coincidence width of the joint spectrum is $\delta\lambda_c = 0.52 \pm 0.01$ nm. According to Ref. [48], the degree of frequency entanglement was determined to be $R = \Delta\lambda_s/\delta\lambda_c = 6.19$. The discrepancy between the experimental result and the theoretical result is mainly caused by the resolution bandwidth of the monochromators. It generated a broaden in the bandwidth of signal and idler photon spectrum as well as the coincidence width of the joint spectrum.

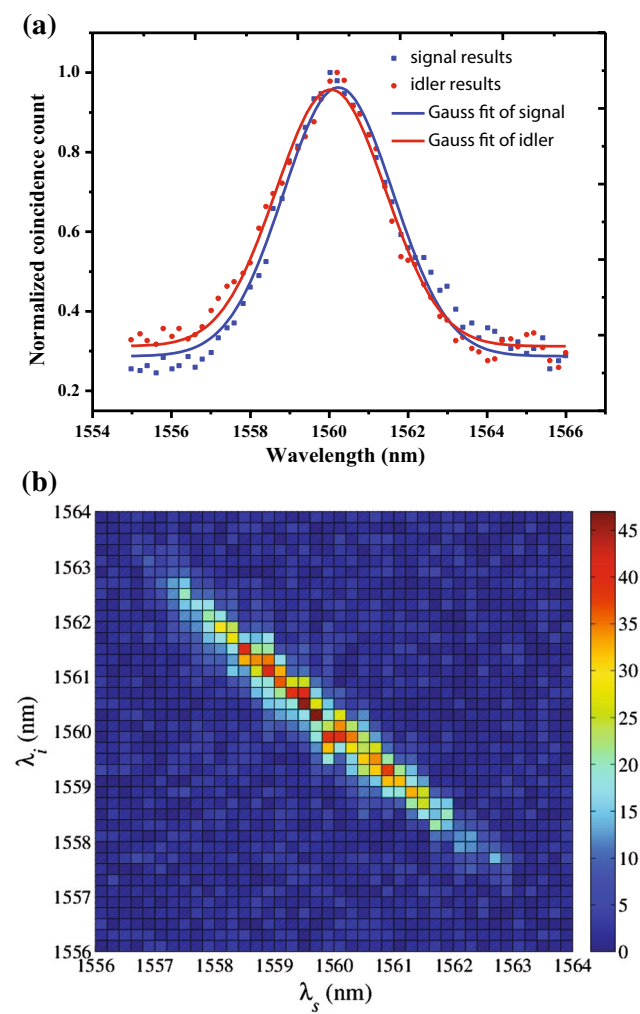

**Fig. 6** The experimental results of the single photon spectra (**a**) and the joint spectrum (**b**)

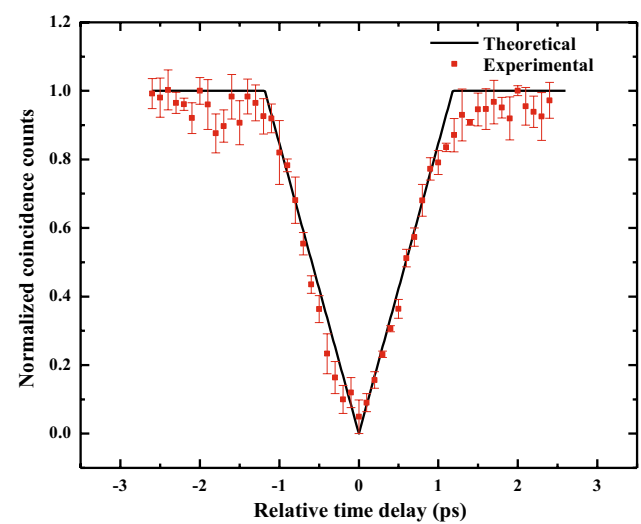

**Fig. 7** HOM interference result. *Black line* is the theoretical HOM interference *curve*, *red dots* are the experimental results

*4.2.2 HOM interference measurement*

After extracting the accidental coincidences and normalizing, the result of the HOM interference measurement is shown by red points in Fig. 7. To estimate the error of visibility, we measured the HOM coincidence distribution for 10 times. Out of the measurements, we got the standard deviation of the visibility results. The HOM interference visibility was measured to be $95 \pm 3\,\%$, with a FWHM coherence time-width of $1.28 \pm 0.02$ ps.

The measurement results have a very good agreement with the theoretical simulations, which shows that the high excess intensity noise of pump laser has no influence on the generated frequency anti-correlated entanglement.

## 5 Conclusions

In summary, we have generated a frequency anti-correlated entangled biphoton source at 1560 nm and experimentally characterized its quantum properties. The results shows that such source has a frequency entanglement degree of 6.19, while the frequency indistinguishability is determined as around $95 \pm 3\,\%$ by HOM interference measurement. The agreement between the measurements and theory have also shown that the high intensity noise of pump laser has no influence on the generated frequency anti-correlated entanglement.

**Acknowledgments** This work is supported by the National Natural Science Foundation of China (11174282, 91336108, 61127901), the Youth Talent Support Plan of the Organization Department of China, the Key Fund of the Western light Talent Cultivation Plan of the CAS, China, the Cross and Cooperative Science and Technology Innovation Team Project of the CAS, China.